\documentstyle[12pt,epsfig,axodraw]{article}
\newcommand{\bi}{\begin{itemize}}
\newcommand{\ei}{\end{itemize}}

\def\CompHEP{C\kern-.3em\lower.17ex\hbox{o}\kern-.135emm\kern-.145emp\kern-.165emH\kern-.125emE\kern-.1emP\ }

\def\PLB{{ Phys. Lett.}  B}
\def\PRL{ Phys. Rev. Lett.}
\def\PRD{{ Phys. Rev.} D}

\def\Journal#1#2#3#4{{#1} {\bf #2}, #3~(#4)}

\begin{document}
\begin {flushright}
FSU-HEP-20010205\\
\end {flushright} 
\vspace{3mm}
\begin{center}
{\Large \bf QCD corrections to FCNC single top production at HERA}
\end{center}
\vspace{2mm}
\begin{center}
{\large A.~Belyaev\footnote{On leave of absence from 
Nuclear Physics Institute, Moscow State University.} 
and Nikolaos Kidonakis\footnote{Present address: Cavendish Laboratory,
University of Cambridge, Cambridge CB3 0HE, England}}\\
\vspace{2mm}
{\it Physics Department\\
Florida State University\\
Tallahassee, FL 32306-4350, USA} \\
\end{center}

\begin{abstract} 
We calculate first-order QCD corrections to the cross section for single top
quark production mediated by flavor changing neutral currents~(FCNC) in $ep$
collisions at the HERA collider. We find that the uncertainty due to 
the choice of the QCD scale is significantly reduced. 
This study is motivated  by the current experimental
work for limits on the  FCNC single top cross section.  

\end{abstract}
\pagebreak

The top quark is the only known fermion which has a mass close to the scale 
of the electroweak symmetry breaking~(EWSB). Hence, the study of the 
electroweak properties of the top-quark sector may shed some light on 
the mechanism responsible for the EWSB. Moreover,  deviations from the  
Standard Model predictions might be
expected in the large mass top sector. Thus, single top quark physics could
probe various  physics beyond the Standard Model: anomalous gluon-top quark
couplings~\cite{glu-top},  anomalous $Wtb$ couplings
~\cite{sngl-wtb},  new strong dynamics~\cite{sngl-strong}, 
flavor changing neutral current
(FCNC) couplings~\cite{sngl-fcnc,effic}, R-parity
violating SUSY effects~\cite{sngl-rpv}, CP-violation
effects~\cite{sngl-cp}, and 
effects of Kaluza-Klein excited $W$-bosons~\cite{sngl-extra}.

Single top quark physics is a very promising place  to test the 
FCNC effects for $tqV$ couplings,
where $q = u$- or $c$-quark and $V=\gamma,Z,g$.  Those couplings
effectively appear in  Supersymmetry or in the scenario where new dynamics
takes place in the fermion mass generation. 
The effective Lagrangian involving such  couplings of a $t,q$ pair 
to massless bosons is the following:
\begin{equation}
\Delta {\cal L}^{eff} =    \frac{1 }{ \Lambda } \,
[ \kappa_{tq\gamma} e   \bar t \sigma_{\mu\nu} q
F^{\mu\nu} + \kappa_{tqg} g_s \bar t
\sigma_{\mu\nu} \frac{\lambda^i}{2} q G^{i \mu\nu}]
+ {\it h.c.}, 
\end{equation}
where $F^{\mu\nu}$ and $G^{i \mu\nu}$ are the usual electromagnetic and gluon
field tensors with  respective FCNC $\kappa_{tq\gamma}$ and $\kappa_{tqg} $
couplings, $\lambda^i$ are the Gell-Mann matrices,
and $\sigma_{\mu \nu}=(i/2)(\gamma_{\mu}\gamma_{\nu}
-\gamma_{\nu}\gamma_{\mu})$ with $\gamma_{\mu}$ the Dirac matrices.
It was found \cite{effic} that the
strength for the anomalous $tcg$
($tug$) coupling may be probed to $\kappa_{tcg} / \Lambda =
0.092\rm{~TeV}^{-1}$~($\kappa_{tug} / \Lambda = 0.026 \rm{~TeV}^{-1}$) at
the Tevatron with $2 \rm{~fb}^{-1}$ of data and $\kappa_{tcg} / \Lambda =
0.013\rm{~TeV}^{-1}$~($\kappa_{tug} / \Lambda = 0.0061 \rm{~TeV}^{-1}$) at
the LHC with $10 \rm{~fb}^{-1}$ of data. Efficiencies from \cite{effic}
can be used to put the limits on $\kappa_{tc\gamma}$, $\kappa_{tu\gamma}$
couplings~($\Lambda=m_{\rm top}$, 95\%CL) at the Tevatron Run 2:
$\kappa_{tc\gamma}<0.24$, $\kappa_{tu\gamma}<0.074$. However, better limits 
(of the order of 0.044 at Run 2) on these couplings $\kappa_{tq\gamma}$ are 
expected to come from  the study of decays $t \rightarrow q \gamma$ of 
pair-produced tops. This
process already allows to derive an upper bound $\kappa_{tq\gamma} < 0.14$ 
from the CDF
data taken at  Tevatron Run 1, which is slightly better than the 
limit obtained by studying $ee \rightarrow tq$ in LEP2 data \cite{LEP2}.

\begin{figure}[htb]
\begin{center}
\begin{picture}(120,120)(0,0)
\Line(0,25)(100,25)\Line(0,75)(100,75) 
\Vertex(50,25){3} \Vertex(50,75){3} 
\Photon(50,25)(50,75){3}{5}
\Text(0,35)[c]{$e$}\Text(0,85)[c]{$u$}
\Text(60,50)[c]{$\gamma$}
\Text(100,35)[c]{$e$}\Text(100,85)[c]{$t$}
\end{picture}
\end{center}
\caption{\label{diag_born}  Diagram for the FCNC singe top quark 
production at HERA}
\end{figure}
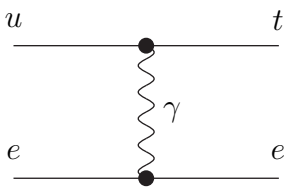

It is interesting to note that HERA should provide  a very good
sensitivity on the FCNC $tu\gamma$ coupling  via single top production
in the process $e u\to e t$ with the respective diagram shown 
in Fig.~\ref{diag_born}.
\vspace*{-1cm}

Even at present ZEUS+H1 160 pb$^{-1}$ integrated luminosity, in the absence 
of a signal, the limit should be $\kappa_{tu\gamma} < \sim 0.05$, which is
significantly better than the current most stringent bound~\cite{liverpool}.
Alternatively, the relatively large~($\sim 1$ pb) 
cross section still allowed by the current 
CDF limit on $\kappa_{tu\gamma}$ would lead to many single
top events. 
It is interesting to note that H1 observed
some  events with a high $p_T$ isolated lepton~($e$ or $\mu$),
together with missing energy and a large $p_T$ hadronic final
state. Such a final state would be expected from single
top events, where the $W$ coming from the top undergoes
a semileptonic decay \cite{H1MUEV}.

However there is a big ambiguity due to the choice of the QCD scale.
FCNC single top quark production at HERA is a $t$-channel process 
involving  a massive final state top quark. The effective QCD scale is unknown:
one can not use $Q^2=-t=(p_e-p_{e'})^2$ as the scale, for example,
since the kinematical distributions concentrated around $Q^2_{min}$
are of the order of $m_e^2\simeq 10^{-7}$ GeV$^2$. This value of $Q^2$
for the parton structure function has no sense. 

At the same time 
$Q^2=m_{\rm top}^2$ seems a little large for this process with such a small
value of the $t$ variable. For the $Q$ scale between $\simeq 5$ GeV 
(minimal $Q$ for the parton density) and $m_{\rm top}=175$ GeV
the Born cross section varies by a factor of 2.
For example, the cross section for a center-of-mass (CM) energy 
$\sqrt{S}=300$ GeV and  
CTEQ5M parton densities~\cite{cteq5}
at  $Q=m_{\rm top}$ is equal to 0.39 pb
and at $Q=5$ GeV equal to 0.78 pb.
This is not surprising for 
the given very different QCD scales.  First-order QCD corrections should 
stabilize the cross section with variation of the QCD scale.
This is the subject of this letter.

We calculate the first-order QCD corrections to the FCNC single top production
using the eikonal approximation. Since the top is very heavy, this
process at HERA is dominated by the threshold region where the eikonal
approximation, which describes soft gluon emission,
is valid. We calculate one-loop eikonal diagrams using techniques
developed for top pair production and other QCD hard scattering cross 
sections \cite{NKGS,KOS,NK}. It is well known that near threshold
the eikonal corrections reproduce the dominant terms of the full 
QCD corrections for a variety of cross sections both analytically and 
numerically \cite{NKGS,NK}. It is also known from resummation studies
that these corrections exponentiate, but we 
will not pursue this topic in this letter.
We define the usual Mandelstam invariants
for the process $e(p_e)+u(p_u) \rightarrow e(p_{e'})+t(p_t)$
as $s=(p_e+p_u)^2$, $t=(p_t-p_u)^2$, and $u=(p_t-p_e)^2$.
We also define a variable $s_2=s+t+u-m_t^2-2m_e^2$, with 
$m_t$ and $m_e$ the top quark and electron masses, respectively. 
At threshold $s_2=0$.
We find that the first-order QCD corrections 
in the $\overline{\rm MS}$ scheme take the form
\begin{eqnarray}
&& \frac{d{\hat \sigma}^{(1)}_{eu\rightarrow et}}{dt \, du}
=F^B_{eu\rightarrow et}
\frac{\alpha_s(\mu_R^2)}{\pi}
\left\{2C_F\left[\frac{\ln(s_2/m_t^2)}{s_2}\right]_{+}\right.
\nonumber \\ && \quad \left.
{}+\left[\frac{1}{s_2}\right]_{+} 
C_F\left[-1-2\ln\left(\frac{-u+m_e^2}{m_t^2}\right)
+2\ln\left(\frac{m_t^2-t}{m_t^2}\right)
-\ln\left(\frac{\mu_F^2}{m_t^2}\right)\right]\right\}
\nonumber \\ && \quad \left. 
{}+\delta(s_2) \left[-\frac{3}{4} 
+\ln\left(\frac{-u+m_e^2}{m_t^2}\right)\right]
C_F\ln\left(\frac{\mu_F^2}{m_t^2}\right) \right\}\, ,
\label{NLO}
\end{eqnarray}
where $\mu_F$ and $\mu_R$ are the factorization and renormalization
scales, respectively, and the Born term, $F^B_{eu\rightarrow et}$, 
is defined by
\begin{equation}
F^B_{eu\rightarrow et}= \frac{\kappa_{\gamma}^2 \, e^4}
{2\pi \, m_t^2 \, t^2 \, (s-m_e^2)^2}
\left\{-t\left[2m_e^4+m_t^4-2s^2+(2s+t)(2s-m_t^2-2m_e^2)\right]
-2m_e^2m_t^4\right\} \, .
\end{equation}
The Born differential cross section is
$d{\hat \sigma}^B_{eu\rightarrow et}/(dt du)=F^B_{eu\rightarrow et} \, 
\delta(s_2)$.
The plus distributions are defined by their integral with any smooth
function $\phi$, such as the parton densities, as \newline
$\int_0^{m^2} ds_2 \, \phi(s_2) \,  [(\ln^n(s_2/m^2))/s_2]_{+} =
\int_0^{m^2} ds_2 \, [(\ln^n(s_2/m^2))/s_2] \, [\phi(s_2) - \phi(0)].$

\begin{figure}[ht]
\noindent
\mbox{\epsfig{file=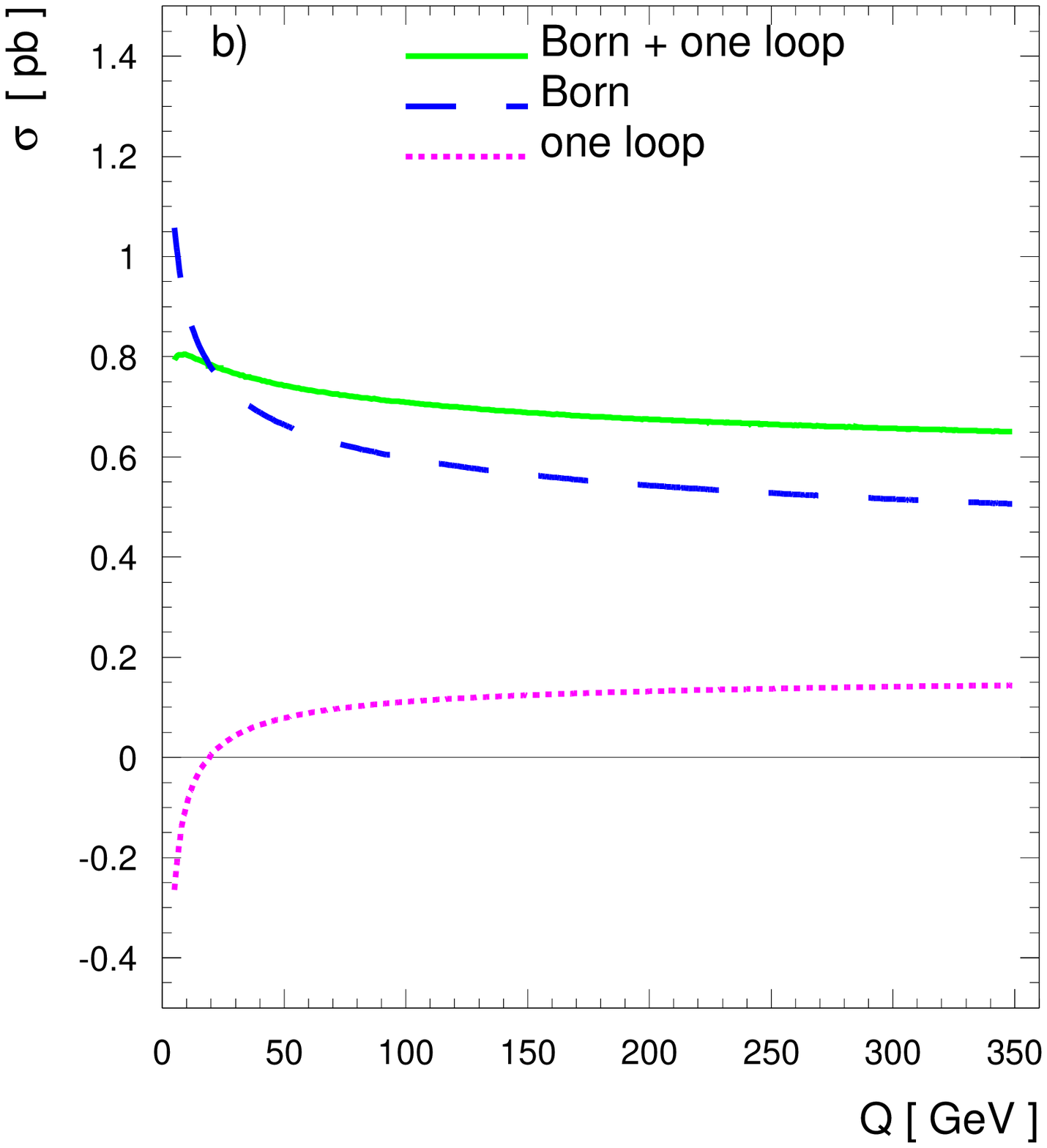,width=0.45\textwidth}}
\mbox{\epsfig{file=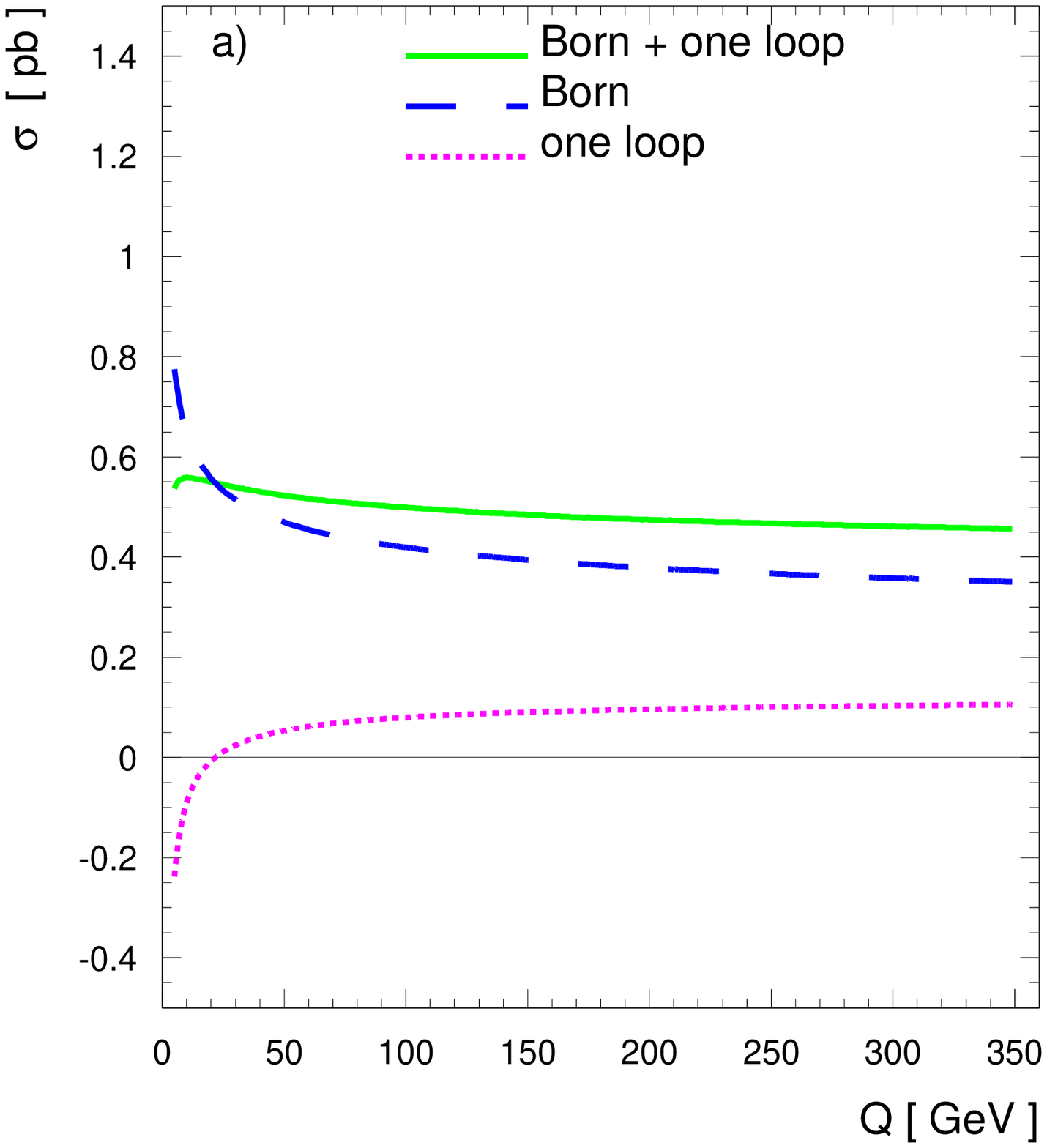  ,width=0.45\textwidth}}
\caption{\label{fig:cs} Born, one loop, and Born+one loop cross section 
for FCNC
single top quark production at HERA with $m_{\rm top}=175$ GeV
and $\kappa_{tu\gamma}=0.1$
for~(a) $\sqrt{S}=300$ and~(b) 318 GeV.}
\end{figure}

\begin{figure}[hb]
\noindent
\mbox{\epsfig{file=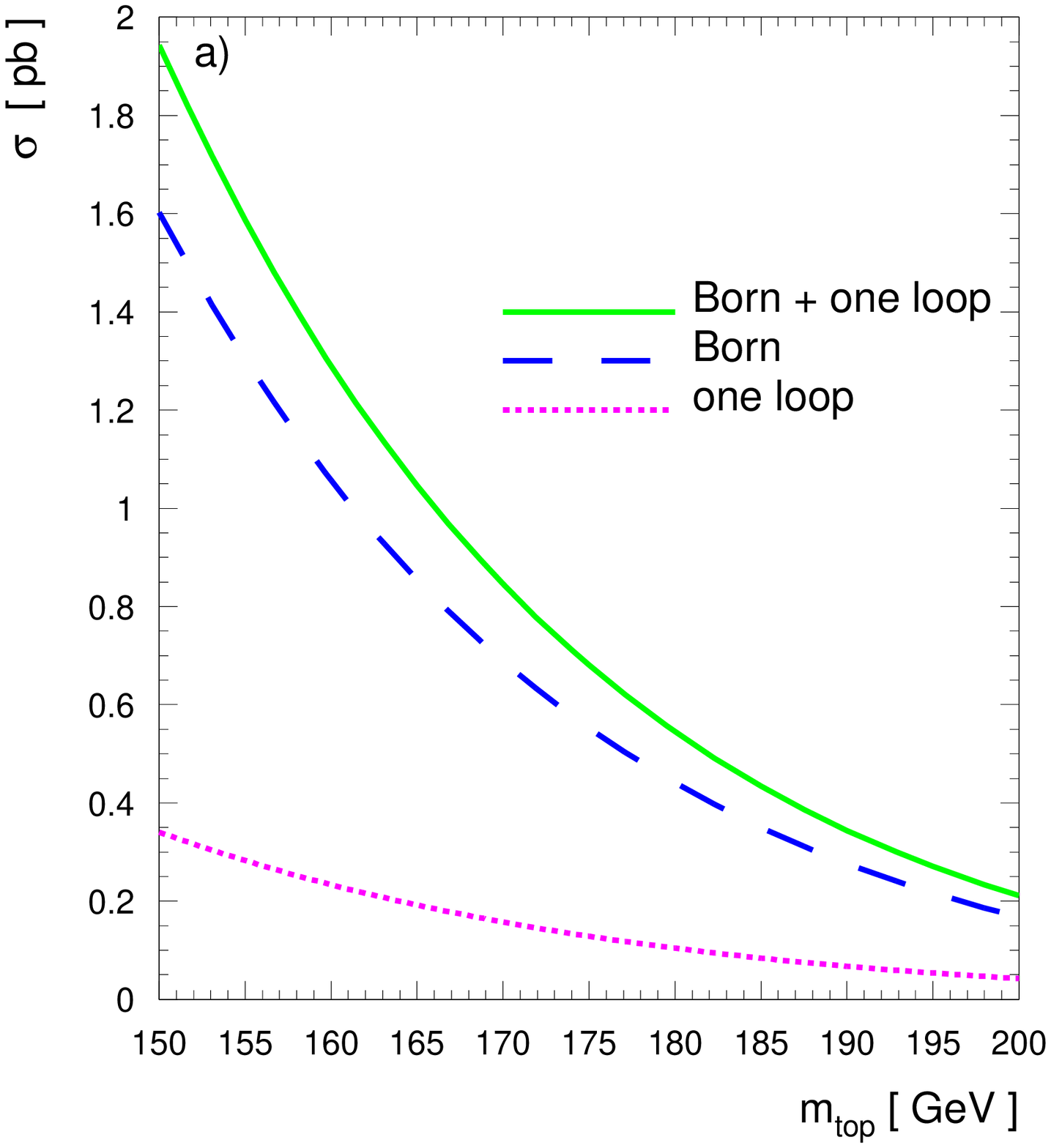,width=0.45\textwidth}}
\mbox{\epsfig{file=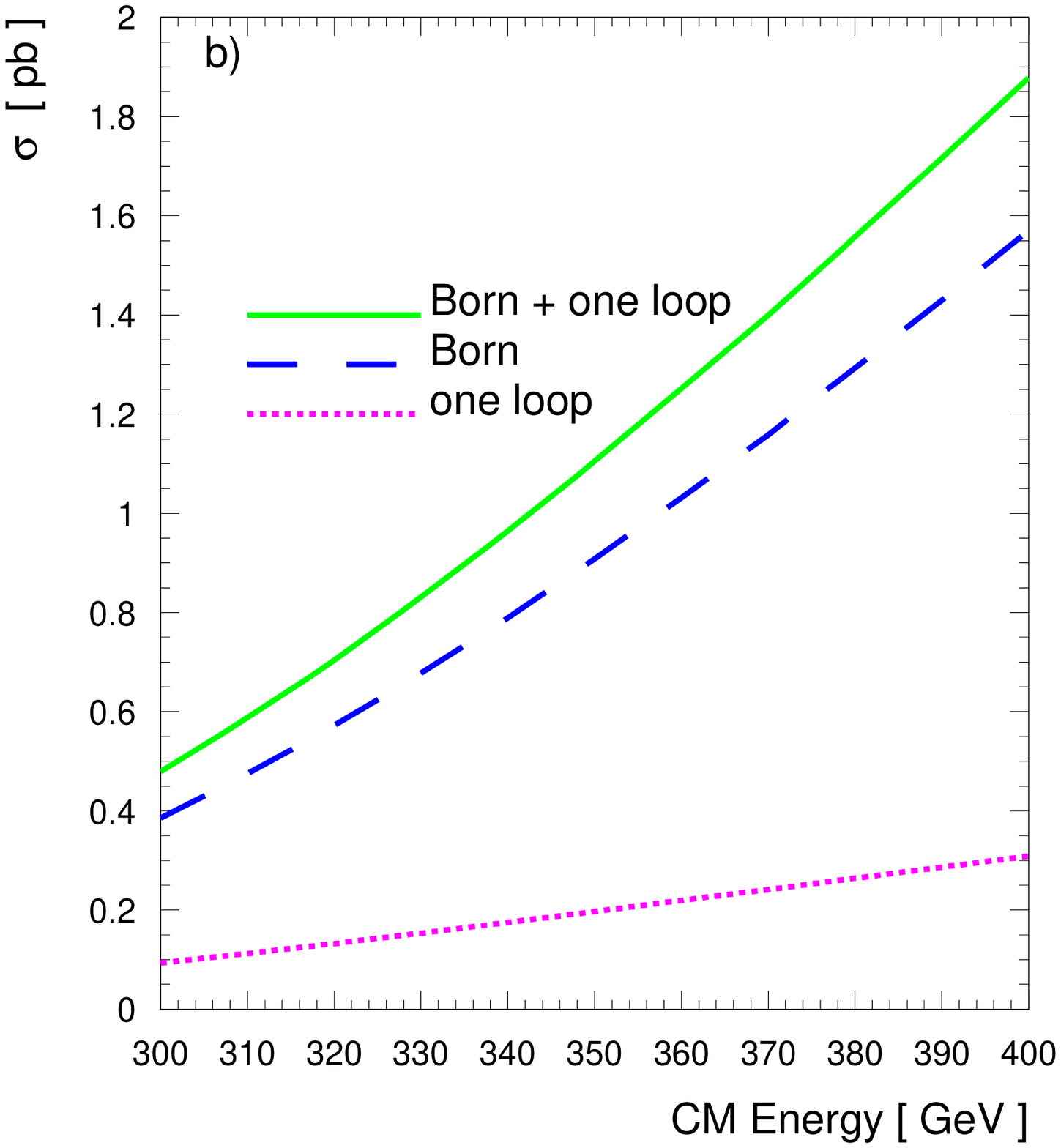  ,width=0.45\textwidth}}
\caption{\label{fig:cs-x} Born, one loop,  and Born+one loop 
cross section for the FCNC
single top quark production at HERA with $Q=m_{\rm top}$
and $\kappa_{tu\gamma}=0.1$ versus 
(a) the top quark mass with $\sqrt{S}=318$ GeV and~(b) the CM energy
with $m_{\rm top}=175$ GeV.}
\end{figure}

In Figs.~\ref{fig:cs}(a),(b) we present the Born and one loop results
for $\sqrt{S}=300$ and 318 GeV, respectively, 
as functions of the QCD scale $Q=\mu_F=\mu_R$.
We note the stabilization of the cross section with scale variation
when the QCD corrections are included.
Since the Born+one loop results are very stable throughout the $Q$ region,
it makes little difference what scale we choose for the central value of $Q$. 
We decided to use $m_{\rm top}$ for the central value and vary the scale
by a factor of two, as commonly done.
We found that for $\sqrt{S}=300$ GeV the QCD corrected cross section is
$\sigma_{eu\to et}^{B+1-{\rm loop}}=0.48 \pm 0.02$ pb
and for $\sqrt{S}=318$ GeV  it is         
$\sigma_{eu\to et}^{B+1-{\rm loop}}=0.68^{+0.04}_{-0.03}$ pb.
The $\kappa_{tu\gamma}$ parameter was chosen to be 0.1.

One should notice that we did not include in our result the uncertainty  due  
to the top quark mass. 
For illustrative purposes we present the Born, one-loop, and 
Born+one loop cross sections 
as functions of the top
quark mass and CM energy in Figs.~\ref{fig:cs-x}(a) and~(b), respectively. 
A variation of the top quark
mass by $\pm 5$ GeV causes over 20\% uncertainty in  the total cross 
section. The role of the
CM energy is very important in this kinematical region, 
where the quark luminosity
quickly  increases with the energy: the 6\% increase of the CM energy 
($300\to 318$ GeV) leads to
a 40\% increase of the total cross section.

The HERA collider 
has the potential to establish the best limit on the FCNC electromagnetic 
coupling involving top and $u$- quarks. Keeping in mind also the observation 
of events with isolated leptons, missing transverse momentum, and a jet 
with large transverse momentum at HERA~(which motivated the experimental 
studies),  an accurate limit on the FCNC coupling based on 
the QCD corrected cross section is very important.

\section*{Acknowledgements}
We would like to thank E. Boos, M. Kuze, J.F. Owens, and E. Perez for useful 
discussions.
This work was supported in part by the U.S. Department of Energy.

\end{document}